\documentclass[aps,superscriptaddress,twoside,twocolumn,floatfix,a4paper,longbibliography, showkeys]{revtex4-1}

\usepackage{graphicx}
\usepackage{amsmath}
\usepackage{dcolumn}   
\usepackage{bm}        
\usepackage{amssymb}   
\usepackage{graphicx,epsfig}
\usepackage{color}
\usepackage[usenames,dvipsnames]{xcolor}
\usepackage{amsmath,amssymb, amsthm, amsfonts}
\usepackage{xspace}
\usepackage{xcolor}
\usepackage{verbatim}
\usepackage{mathtools}
\usepackage{lipsum}
\usepackage{natbib}
\usepackage{inputenc}

\begin{document}
\title{Development of highly sensitive nanoscale transition edge sensors for gigahertz astronomy and dark matter search}

\author{Federico Paolucci}
\email{federico.paolucci@nano.cnr.it}
\affiliation{NEST, Istituto Nanoscienze-CNR and Scuola Normale Superiore, I-56127 Pisa, Italy}
\affiliation{INFN Sezione di Pisa, Largo Bruno Pontecorvo, 3, I-56127 Pisa, Italy}

\author{Vittorio Buccheri}
\affiliation{INFN Sezione di Pisa, Largo Bruno Pontecorvo, 3, I-56127 Pisa, Italy}
\affiliation{NEST, Istituto Nanoscienze-CNR and Scuola Normale Superiore, I-56127 Pisa, Italy}

\author{Gaia Germanese}
\affiliation{NEST, Istituto Nanoscienze-CNR and Scuola Normale Superiore, I-56127 Pisa, Italy}
\affiliation{Dipartimento di Fisica dell'Universit\`a di Pisa, Largo Pontecorvo 3, I-56127 Pisa, Italy}

\author{Nadia Ligato}
\affiliation{NEST, Istituto Nanoscienze-CNR and Scuola Normale Superiore, I-56127 Pisa, Italy}

\author{Riccardo Paoletti}
\affiliation{Dipartimento di Scienze Fisiche, della Terra e dell'Ambiente dell'Universit\`a di Siena, Strada Laterina, 8 I-53100 Siena, Italy}
\affiliation{INFN Sezione di Pisa, Largo Bruno Pontecorvo, 3, I-56127 Pisa, Italy}

\author{Giovanni Signorelli}
\affiliation{INFN Sezione di Pisa, Largo Bruno Pontecorvo, 3, I-56127 Pisa, Italy}

\author{Massimiliano Bitossi}
\affiliation{INFN Sezione di Pisa, Largo Bruno Pontecorvo, 3, I-56127 Pisa, Italy}

\author{Paolo Spagnolo}
\affiliation{INFN Sezione di Pisa, Largo Bruno Pontecorvo, 3, I-56127 Pisa, Italy}

\author{Paolo Falferi}
\affiliation{IFN-CNR and Fondazione Bruno Kessler, via alla Cascata 56, I-38123 Povo, Trento, Italy}
\affiliation{INFN, TIFPA, via Sommarive 14, I-38123 Povo, Trento, Italy}

\author{Mauro Rajteri}
\affiliation{Istituto Nazionale di Ricerca Metrologica (INRIM), Str. delle Cacce, 91, I-10135 Torino, Italy }

\author{Claudio Gatti}
\affiliation{INFN, Laboratori Nazionali di Frascati, Via Enrico Fermi, 54, I-00044 Frascati, Italy}

\author{Francesco Giazotto}
\email{francesco.giazotto@sns.it}
\affiliation{NEST, Istituto Nanoscienze-CNR and Scuola Normale Superiore, I-56127 Pisa, Italy}


\begin{abstract}
  Terahertz and sub-terahertz band detection has a key role both in fundamental interactions physics and technological applications, such as medical imaging, industrial quality control and homeland security. In particular, transition edge sensors (TESs) and kinetic inductance detectors (KIDs) are the most employed bolometers and calorimeters in the THz and sub-THz band for astrophysics and astroparticles research. 
  Here, we present the electronic, thermal and spectral characterization of an aluminum/copper bilayer sensing structure that, thanks to its thermal properties and a simple miniaturized design, could be considered a perfect candidate to realize an extremely sensitive class of nanoscale TES (nano-TES) for the giga-therahertz band. Indeed, thanks to the reduced dimensionality of the active region and the efficient Andreev mirror (AM) heat confinement, our devices are predicted to reach state-of-the-art TES performance. In particular, as a bolometer the nano-TES is expected to have a noise equivalent power (NEP) of $5\times10^{-20}$ W/$\sqrt{\mathrm{Hz}}$ and a relaxation time of $\sim 10$ ns for the sub-THz band, typical of cosmic microwave background studies. When operated as single-photon sensor, the devices are expected to show a remarkable frequency resolution of 100 GHz, pointing towards the necessary energy sensitivity requested in laboratory axion search experiments. Finally, different multiplexing schemes are proposed and sized for imaging applications.        
\end{abstract}

\keywords{transition edge sensor, nanoscale, superconductivity, Andreev mirrors, gigahertz, axion, cosmic microwave background}

\maketitle

\section{Introduction}
\label{intro}
In the last decade, astronomy and astrophysics have broadened their interest towards low energy phenomena, such as cosmic microwave background (CMB) \cite{Sironi}, atomic vibrations in galaxy clusters \cite{Villaescusa-Navarro}, and new particles in dark matter \cite{Redondo}. To obtain physical insight of these phenomena, the detection of faint signals in the micro- (terahertz) and sub-millimeter (gigahertz) spectral range plays a fundamental role. To this end, the key ingredient is the development of new ultrasensitive bolometers and single-photon detectors, i.e., calorimeters. On the one hand, the temperature and polarization maps of the CMB fluctuations \cite{Seljak,Kamionkowski} and the detection of polarized radiation due to the hydrogen atom emission in the galaxy clusters \cite{Armus} are the main astronomy applications for gigahertz (GHz) and terahertz (THz) bolometers. On the other hand, low frequency calorimeters could play a fundamental role in axions search \cite{Ringwald}, one of the principal candidates for the dark matter. Axions are very weakly interacting particles with small mass ($\sim 1$~meV) thus impossible to be revealed by means of colliders. Therefore,  light-shining-through-wall (LSW) experiments have been proposed to generate axion-like particles (ALP) in the laboratory \cite{Spagnolo}, differently from experiments focused on space surveys, such as CAST \cite{Arik} and IAXO \cite{Armengaud}.

Nowadays, the most employed detectors in the THz energy band are the superconducting sensors, such as transition edge sensors (TESs) \citep{Irwin1995a, Irwin2006, Karasik} and kinetic inductance detectors (KIDs) \cite{Monfardini}, for their high sensitivity, robustness and mature technology. The state-of-the-art of these detectors in bolometric operation shows a noise equivalent power (NEP) of $\sim 10^{-19}$ $\mathrm{W/\sqrt{Hz}}$ ~for TESs \cite{Khosropanah} with large active area ($\sim$ 100 $\mathrm{\mu}\mathrm{m^2}$), and $\sim 10^{-18}$ $\mathrm{W/\sqrt{Hz}}$ for KIDs \cite{Visser}. More sensitive and efficient superconducting detectors have been proposed and realized by taking advantage of device miniaturization \cite{Wei} and Josephson effect. For instance, detectors based on superconductor/normal metal/superconductor (SNS) junctions showed a \textit{NEP} of the order of $10^{-20}$ $\mathrm{W/\sqrt{Hz}}$ \cite{Kokkoniemi}, cold electron bolometers showed a $NEP\sim 3\times 10^{-18}$ $\mathrm{W/\sqrt{Hz}}$ \cite{kuzmin2019}, devices based on the temperature-to-phase conversion (TPC) are expected to provide $NEP\sim 10^{-23}$ $\mathrm{W/\sqrt{Hz}}$ \cite{Virtanen}, while a fully superconducting tunable Josephson escape sensor (JES) showed a record intrinsic \textit{NEP} as low as $10^{-25}$ $\mathrm{W/\sqrt{Hz}}$ \cite{Paolucci}.  

To push the TES technology towards lower values of \textit{NEP} with the possibility to detect single photons in the GHz band \cite{Alesini}, a strong reduction of the thermal exchange mechanisms of the active region, i.e., the portion of the device transitioning to the normal-state when radiation is absorbed, is necessary \cite{Bergmann}. To this end, we envision a nanoscale TES (nano-TES) exploiting a simple and sturdy miniaturized design together with the Andreev mirrors (AM) effect \cite{andreev} to thermally isolate the sensor active region. 
Here, we present and experimentally characterize the active region of the nano-TES structures by an electrical and a thermal points of view. For simplicity, we will adopt the notation nano-TES to indicate the structures analyzed in this paper.
Indeed, operating in the bolometer configuration our nano-TESs would reach a total noise equivalent power ($NEP_{tot}$) of $\sim~10^{-20}$ $\mathrm{W/\sqrt{Hz}}$, while as calorimeters they are expected to reach a resolving power ($\delta \nu/\nu$) of $10$ in sub-THz band. Their measured thermal and electrical performance are several orders of magnitude better than devices with identical dimensions but without the Andreev mirrors heat constrictions. Finally, we propose and size two possible multiplexing circuits in frequency domain (FDM) and microwave resonators (MR) as readout for a nano-TES array enabling the realization of multi-pixel cameras. In addition to gigahertz astronomy and particle physics, the nano-TES could find applications for medical imaging \cite{Sun}, industrial quality controls \cite{Ellrich} and security \cite{Rogalski}.

This paper is organized as follows. Section \ref{sec_devicefabrication}  reports the simple fabrication of the nano-TES and of a secondary device used to extract all the parameters of the active region. Section \ref{sec_nanotesproperties} describes the electrical properties of the nano-TES. The spectral and thermal characterization of the active region are resumed in Sec. \ref{sec_spectral} and \ref{sec_thermal}, respectively. The expected nano-TES performance as a bolometer and a calorimeter deduced from the experimental data are reported in Sec. \ref{Sec:Performance} together with the comparison of a TES of identical materials and dimensions but not equipped with AM. Finally, Sec. \ref{Sec:Readouts} covers possible multiplexing readout circuits to design multi-pixel cameras.

\begin{figure}[htbp]
    \includegraphics[width=0.48\textwidth]{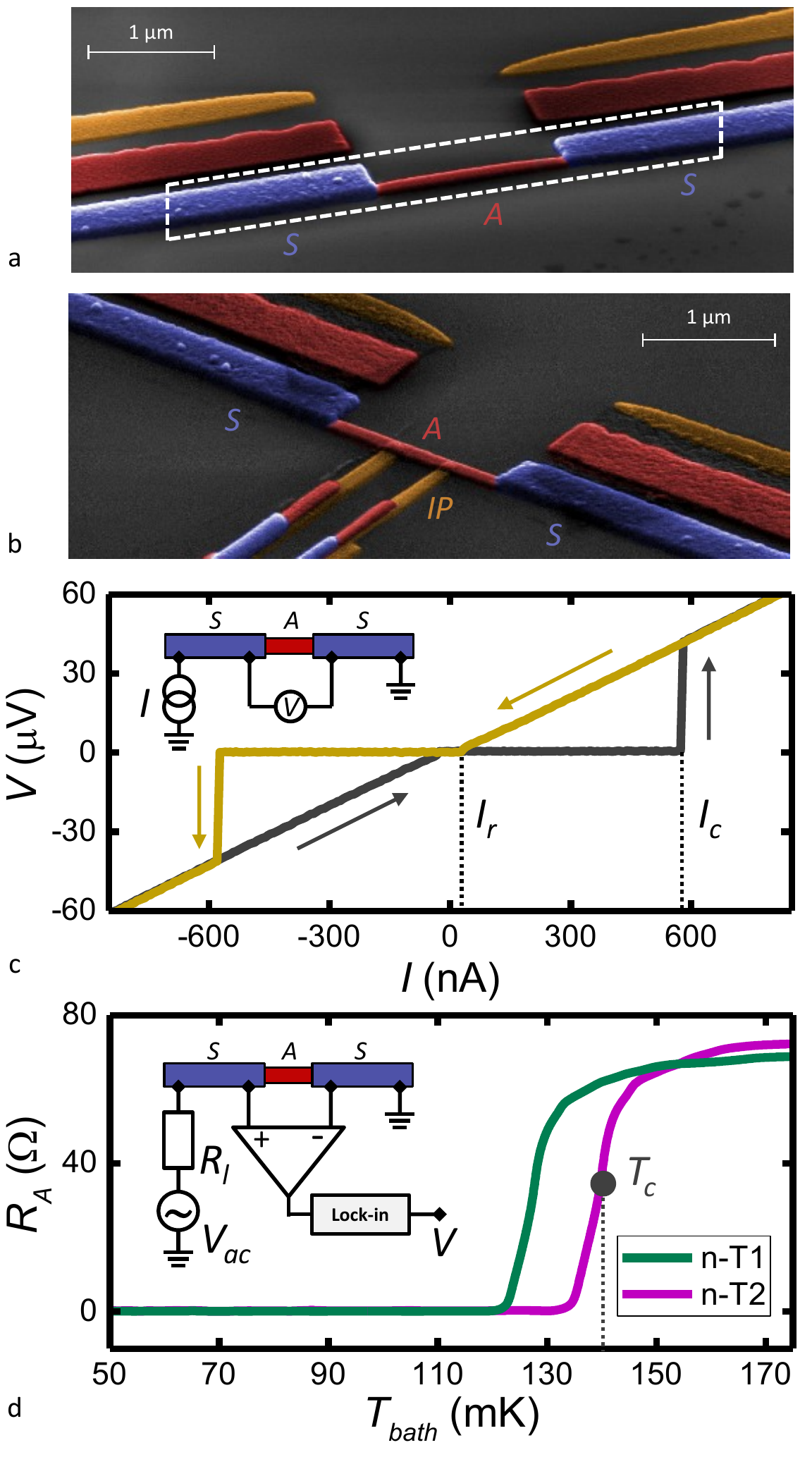}
    \caption{False-color SEM pictures of the nano-TES (a) and SD (b). The three different layers are the active region Al/Cu bilayer $A$ (red), the Al electrodes $S$ (blue) and the Al-oxidized tunnel probes $IP$ (yellow). The detector structure is pointed out by the dashed white box in the nano-TES SEM. (c) Four-terminal $VI$ measurement of sample n-T1 for a positive (grey line) and a negative (yellow line) current slope at $T_{bath}=20$ mK. The black dotted lines highlighted the critical current $I_{c}\sim$ 600 nA and the retrapping $I_{r}\sim$ 27 nA of $A$, where the  $\mathcal{S}\rightarrow\mathcal{N}$ and the $\mathcal{N}\rightarrow\mathcal{S}$ transitions occur, respectively. \textit{Inset}: Experimental setup for the $VI$ measurement. (d) Temperature dependence of the nano-TES active region resistance, $R_{A}$, for n-T1 (green line) and n-T2 (purple line). By following a positive $T_{bath}$ slope, $R_{A}$ goes from zero to the normal state resistance value. The critical temperature $T_{c}$ (grey point for n-T2) is the temperature corresponding to the half of the normal resistance of the samples. \textit{Inset}: Experimental setup for the $R_{A}$ versus $T_{bath}$ measurements.}
    \label{Fig_1}
\end{figure}

\section{Devices fabrication}
\label{sec_devicefabrication}

The experiments discussed in this paper are performed thanks to two different device architectures: the nano-TES and a secondary device (SD). Measurements on the nano-TES provided the active region resistance versus temperature characterization, $R_{A}(T)$, the active region critical $I_{c}$ and retrapping $I_{r}$ current, and the critical temperature $T_{c}$. Instead, the active region spectral and thermal properties have been obtained by the experiments performed on the SD.

The false-color scanning electron microscope (SEM) pictures of a typical nano-TES and SD are shown in Fig. \ref{Fig_1}-a and -b, respectively. The nano-TES, highlighted by the dashed white box in Fig. \ref{Fig_1}-a, consists in a 1.5 $\mu$m-long, 100 nm-wide and 25 nm-thick Al/Cu bilayer nanowire-like active region (red), which is sandwiched between the Al electrodes (blue). Since the superconducting gap of the Al layer is higher than that of the Al/Cu bilayer (due to inverse proximity effect \cite{tinkham}), the electrodes act as AM for the active region, leading to the advantages discussed in Sec. \ref{sec_thermal}. The same nano-TES structure is visible in the SD (see Fig. \ref{Fig_1}-b) with the addition of two oxidized Al probes (yellow) lying under the active region forming two tunnel Josephson junctions (JJs) \cite{Giazotto}. These two Al probes allow to characterize both the energy gap and the thermal properties of the active region.

Both the nano-TES and the SD were realised during the same fabrication process, ensuring the homogeneity of their properties. They were fabricated by electron-beam lithography (EBL) and 3-angles shadow mask evaporation of metals onto a silicon wafer covered with 300 nm of thermally grown SiO$_{2}$. The evaporation was performed in an ultra-high vacuum electron-beam evaporator with base pressure of about $10^{-11}$ Torr. 
By referring to the color code of Fig \ref{Fig_1}-a and -b, the fist Al layer (yellow) with thickness 13 nm was evaporated at an angle of -40$^\circ$ and then oxidized by exposition to 200 mTorr of O$_{2}$ for 5 minutes to obtain the tunnel probes in the SD. In a second step, the Al/Cu bilayer (red) was evaporated at an angle of $0^\circ$ to form the active region with partial thicknesses $h_{Al} = 10.5$ nm and $h_{Cu} = 15$ nm for the aluminum and copper layer, respectively. Finally, a second Al layer (blue) of thickness 40 nm was evaporated at an angle of $+40^\circ$ to obtain the AM electrodes.

The notation \textit{A}, \textit{S}, \textit{P} and \textit{I} will be used to indicate the Al/Cu active region, the Al electrodes, the Al probes and the probes insulating barrier, respectively. The nano-TES measurements have been performed on two different devices, n-T1 and n-T2. All the following experiments have been performed in a $^{3}$He-$^{4}$He dilution refrigerator with bath temperature $T_{bath}$ ranging from 20 mK to 250 mK.

\section{Nano-TES electrical properties}
\label{sec_nanotesproperties}

The four-terminal voltage-current $VI$ characteristics of n-T1 are shown in Fig. \ref{Fig_1}-c at $T_{bath}=20$ mK for a positive (grey line) and a negative (yellow line) current slope. The electrical measurement setup is schematized in the inset. Here, the normal $\mathcal{N}$ and the superconducting $\mathcal{S}$ state are recognizable by the linearly growing and the flat behaviour of the $VI$ traces, respectively. The $\mathcal{S}\rightarrow\mathcal{N}$ transition occurs at the critical current $I_{c}\sim 600$ nA, whereas the $\mathcal{N}\rightarrow\mathcal{S}$ occurs at the retrapping current $I_{r}\sim$ 27 nA \cite{courtois}. The normal-state resistance $R_A$ is obtained by the slope of the $VI$ and gets value $R_{A,n-T1}=70$ $\Omega$, while it is obviously zero in the $\mathcal{S}$. This change of $R_{A}$ has a key role in the operation mechanism of a TES detector \cite{Irwin1995a} (see Sec. \ref{Sec:Readouts} for further details).

The temperature dependence of $R_{A}$ is shown in Fig. \ref{Fig_1}-d for n-T1 (green line) and n-T2 (purple line). The inset shows the experimental setup. The AC current bias is produced by applying a voltage $V_{ac}$ at 13.33 Hz to a load resistance $R_{l}=100$ k$\Omega$ ($R_{l}>>R_{A}$) in order to obtain $I_{ac}=15$ nA independent from $R_A$. The voltage drop $V$ across the nano-TES is measured as a function of $T_{bath}$ via a voltage pre-amplifier connected to a lock-in amplifier. By rising $T_{bath}$, the resistance changes from zero to its normal-state value, by following an edge transition behaviour. The nano-TES critical temperature $T_{c}$ is defined as the temperature corresponding to half of the normal-state resistance (grey point for n-T2 in Fig. \ref{Fig_1}-d). Thus, we have $T_{c1}= 128$ mK and $T_{c2}= 139$ for n-T1 and n-T2, respectively.

 \begin{figure}[t!]
     \centering
     \includegraphics[width=0.5\textwidth]{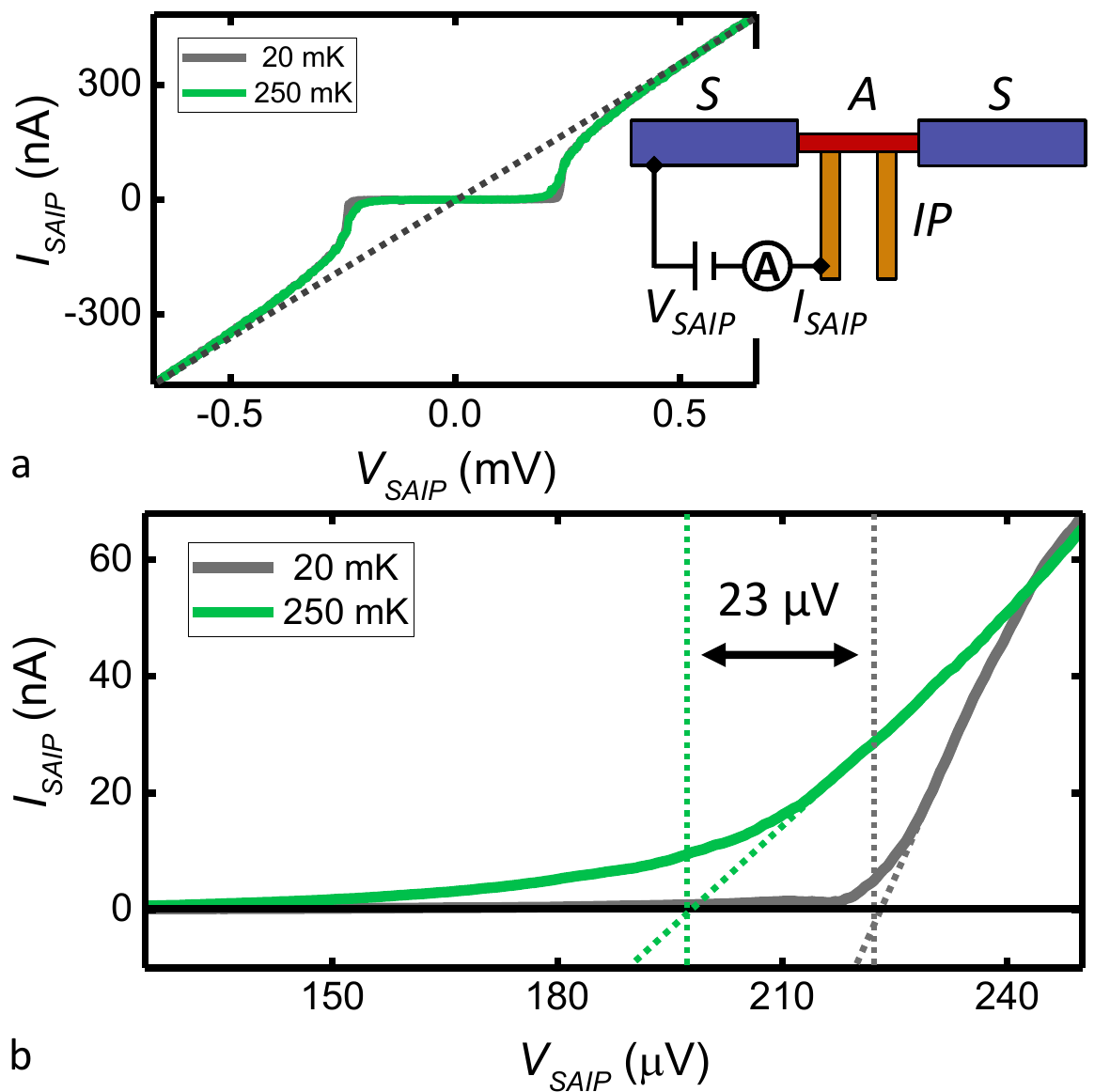}
     \caption{(a) The $IV$ characteristic of the \textit{SAIP} junction at $T_{bath}=20$ mK (grey line) and $T_{bath}=250$ mK (green line). The slope of the linear region (dotted line) represents the $\mathcal{N}$-state tunnel resistance of the JJ, $R_{I}\simeq 12$ k$\Omega$. \textit{Inset}: Schematic representation of SD with the experimental setup used for the spectral characterization. (b) The $IV$ characteristics of the \textit{SAIP} junction zoomed on the positive switching point for two bath temperatures: the high temperature switching point corresponds to $\Delta_{0,P}\simeq200\;\mu$eV, while the difference between the two onset points is $\Delta_{0,A}\simeq23\;\mu$eV.
     }
     \label{fig_spectral}
 \end{figure}
 
\section{Spectral characterization of the active region}
\label{sec_spectral} 

For $T<0.4T_c$, the energy gap of a superconductor [$\Delta(T)$] is temperature independent and equals its zero-temperature value $\Delta_0$ \cite{tinkham}. For higher values of temperature, $\Delta(T)$ decreases monotonically and finally disappears at $T_c$.
Typically, aluminum thin films show a $T_c$ higher than the bulk Al value ($\sim1.2$ K) \cite{cochran}. Therefore, the superconducting gaps of \textit{S} and \textit{P}, $\Delta_{S} (T)$ and $\Delta_{P}(T)$, are temperature independent up to at least 500 mK, thus preserving their zero-temperature values ($\Delta_{0,S}$ and $\Delta_{0,P}$). By contrast, due to inverse proximity effect \cite{tinkham}, superconductivity in $A$ is strongly suppressed. In fact, our resistance versus temperature experiments showed a value $T_{c,A}\simeq 140\ll500$ mK (see Fig. \ref{Fig_1}-d), thus enabling the possibility to independently determine both $\Delta_{0,A}$ and $\Delta_{0,P}$.

To this end, the $IV$ characteristics of a \textit{SAIP} JJ were measured at base temperature and just above $T_{c,A}$, as reported in Fig. \ref{fig_spectral}-a with the grey and the green line, respectively. The experimental setup of the measurements is schematically shown in the inset of the panel. At base temperature, the JJ switches to the $\mathcal{N}$-state when the voltage bias reaches $V_{SAIP}=\pm(\Delta_{0,A}+\Delta_{0,P})/e$ \cite{Giazotto}, where $e$ is the elementary charge. Instead, at $T_{bath}= 250$ mK the transition occurs at $V_{SAIP}=\pm\Delta_{0,P}/e$, since \textit{A} is in the $\mathcal{N}$-state. The tunnel resistance of the JJ is given by the slope of the $IV$ characteristic in the linear region, as highlighted by the black dotted line, and gets value $R_I\simeq 12$ k$\Omega$. 

In order to provide a precise evaluation of the energy gaps, the $IV$ characteristics are zoomed around the switching points acquired for positive voltage bias, as shown in Fig. \ref{fig_spectral}-b. 
The measurement at $T_{bath}= 250$ mK (green line) indicates a value of the zero-temperature superconducting gap of the aluminum probes $\Delta_{0,P}\simeq200\;\mu$eV, corresponding to a critical temperature $T_{c,P}=\Delta_{0,P}/(1.764k_B)\simeq 1.3$ K.
Note that $\Delta_P\sim\Delta_S$, since electrodes and probes thickness are similar \cite{meservey}, thus $T_{c,S}\sim1.3$ K as well.
Instead, the difference between the results obtained at 20 mK and 250 mK leads to $\Delta_{0,A}\simeq23\;\mu$eV corresponding to a critical temperature $T_{c,A}\simeq150$ mK, in good agreement with the data reported in Fig. \ref{Fig_1}-d. 

Importantly, the superconducting gap of \textit{A} is constant along the out-of-plane axis (i.e., the sample thickness), because the bilayer is within the Cooper limit \cite{DeGennes1964,Kogan1982}. In fact, the aluminum thin film follows $h_{Al}=10.5\;\text{nm}\ll\xi_{0,Al}=\sqrt{\hbar D_{Al}/\Delta_{0,Al}}\simeq80$ nm (where $D_{Al}=2.25\times 10^{-3}$ m$^2$s$^{-1}$ is the diffusion constant of Al and $\Delta_{0,Al}\simeq200\;\mu$eV is the superconducting energy gap), while the copper layer respects $h_{Cu}=15\;\text{nm}\ll\xi_{0,Cu}=\sqrt{\hbar D_{Cu}/(2\pi k_B T)} \simeq255$ nm (where $D_{Cu}=8\times 10^{-3}$ m$^2$s$^{-1}$ is the copper diffusion constant and the temperature is chosen $T=150$ mK thus higher than the nano-TES operation value). Furthermore, the active region is much thinner than its superconducting coherence length, that is $h_A=h_{Al}+h_{Cu}=25.5\;\text{nm}\ll\xi_A=\sqrt{l \hbar/[(h_{Al} N_{Al}+h_{Cu}N_{Cu}) R_Ae^2\Delta_{0,A}]}\simeq 220$ nm, where $e$ is the electron charge, while $N_{Al}=2.15\times 10^{47}$ J$^{-1}$m$^{-3}$ and $N_{Cu}=1.56\times 10^{47}$ J$^{-1}$m$^{-3}$ are the density of states ($DOS$s) at the Fermi level of aluminum and copper, respectively. 
 

 \begin{figure*}[htbp]
     \centering
     \includegraphics[width=1\textwidth]{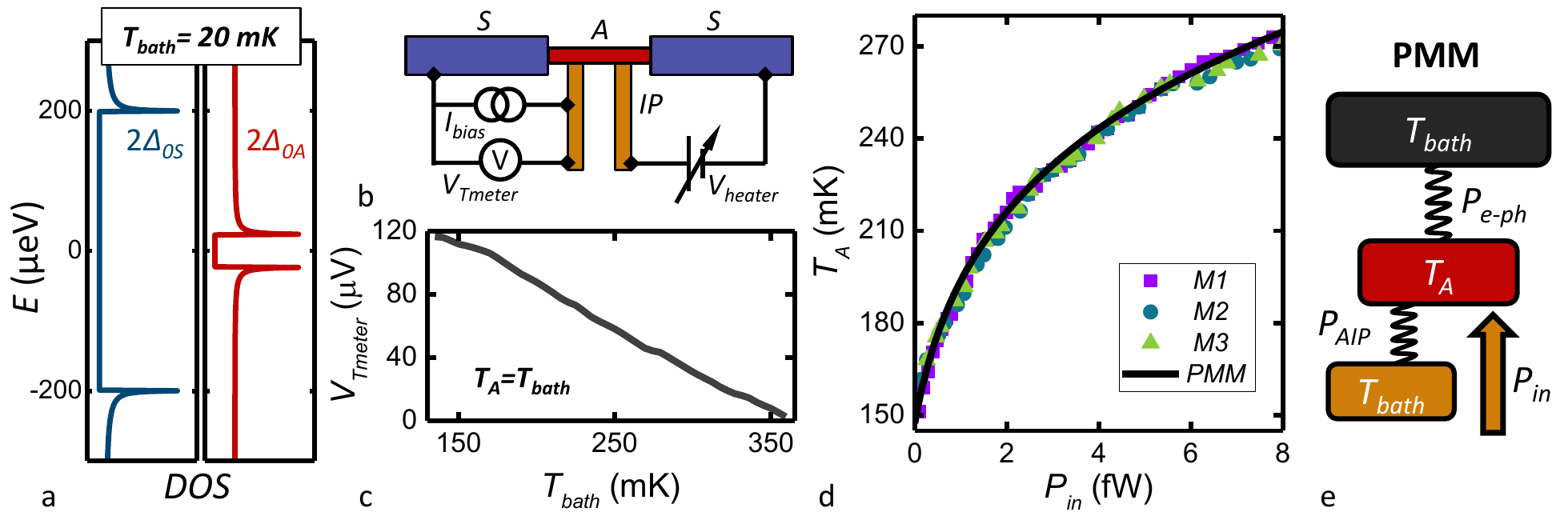}
     \caption{(a) BCS $DOS$s of \textit{S} (blue) and \textit{A} (red) at $T_{bath}=20$ mK with the gaps obtained experimentally, $\Delta_{A}=23\;\mu$eV and $\Delta_{S}=200\;\mu$eV. (b) Schematic representation of the experimental setup used for the thermal characterization: the left and the right \textit{SAIP} JJ are used as electron thermometer and heater, respectively. 
     (c) Thermometer calibration curve which links the voltage output $V_{Tmeter}$ to the $A$ electronic temperature $T_{A}$.
     (d) Electronic temperature of $A$ as a function of the input power for three different data sets \textit{Mi}, with $i=1,2,3$, (colored symbols) at $T_{bath}= 150$ mK. The fitting curve (black line) is obtained by solving Eq. \eqref{balance_equation} of the PMM.
     (e) PMM thermal model of \textit{A}: the input power from the heater $P_{in}$ relaxes through the outward components: $P_{e-ph}$, due to the electron-phonon interaction, and $P_{AIP}$, due to the losses through the thermometer tunnel junction.}
     \label{fig_thermal1}
 \end{figure*}
 
\section{Thermal characterization of the active region}
\label{sec_thermal}

Energy exchange has a key role in determining the nano-TES performance, such as sensitivity and response time, since the increase of the $A$ electronic temperature $T_{A}$ due to the incident radiation strongly depends on the capability of maximizing the thermal confinement. The scope of this section is to study the most prominent heat exchange mechanisms in the active region of the nano-TES for typical operating conditions.

Metallic elements in mesoscopic devices at sub-kelvin temperatures show weak coupling between the electron and the phonon thermal subsystems \cite{Giazotto}, which can lead to $T_{e}\ne T_{ph}$, where $T_{e}$ and $T_{ph}$ are the electron and phonon temperature, respectively. Due to the thickness of the films lower than the phonon wavelength and vanishing Kapitza resistance, the device phonons are thermally anchored to the substrate ($T_{ph}=T_{sub}$) \cite{wellstood}, so that the temperature of both systems can be considered as a parameter set via the refrigerator temperature $T_{bath}$. The geometry of our device also guarantees electronic temperature of the superconducting electrodes $T_{S}$ and the tunnel probes $T_{P}$ equal to the phonon temperature, that is $T_{S}=T_{P}=T_{bath}$. By contrast, the $A$ electronic temperature $T_{A}$ is the fundamental thermal variable in the nano-TES operation mechanism. In general, the value of $T_{A}$ results from the balance between the main thermal exchange channels of \textit{A}. In our case: 
\begin{equation}
P_{in}= P_{e-ph}+P_{AIP}+P_{loss},
\label{balance_equation_2}
\end{equation}
where $P_{in}$ is the power injected, $P_{e-ph}$ is the electron-phonon relaxation, $P_{loss}$ represents the heat losses through $S$ and $P_{AIP}$ is the energy diffusion by an $IP$ probe. Note that the electron-photon interaction contribution has not been considered in the model, since it is negligibly small with comparison to the other thermal channels \cite{bosisio}.

The use of $S$ with energy gap much larger than $A$ can ensure negligible heat out-diffusion from $A$ to $S$. Indeed, the normalised \textit{DOS} of a superconductor reads \cite{Giazotto}:
\begin{equation}
DOS(E, T)=\frac{|E|}{\sqrt{E^2-\Delta^2(T)}}\Theta(E^2-\Delta^2(T)).
\label{eq_dos}
\end{equation}
Thus, the zero-temperature energy-dependent \textit{DOS} of \textit{S} and \textit{A} are calculated by inserting the measured values of $\Delta_{0,S}$ and $\Delta_{0,A}$. The resulting functions are shown in Fig. \ref{fig_thermal1}-a with the blue and the red line, respectively. The thermally excited quasi-particles in \textit{A} do not find available states towards \textit{S}, thus the resistance for heat diffusion exponentially rises by decreasing the bath temperature \cite{andreev}. In particular, at $k_BT_A\ll\Delta_{S}$ the superconducting leads act as AM, namely as perfect barriers for energy diffusion ($P_{loss}=0$). In addition, the big difference between the two superconducting gap ensures that the nano-TES superconducting to dissipative transition affects only $A$, leading to a better control in the resistance change and a small overheating of the detector.

The experimental setup employed to perform the thermal study of \textit{A} is schematically shown in Fig. \ref{fig_thermal1}-b: the left \textit{SAIP} junction was current-biased (at $I_{bias}$) to operate as thermometer, whereas the right JJ was voltage-biased (at $V_{heater}$) to work as heater \cite{Giazotto}. The thermometer has been calibrated by varying $T_{bath}$ and measuring $V_{Tmeter}$ at $I_{bias}= 10$ pA and $V_{heater}=0$ V, as reported in Fig. \ref{fig_thermal1}-c. The bath temperature ranges from  $\sim150$ mK to $\sim350$ mK, so $\Delta_{A}=0$, i.e. \textit{A} is in the normal state, whereas $\Delta_{S}\simeq 200\;\mu eV$. In this normal metal ($A$)/insulator/superconductor ($P$) JJ, the $IV$ characteristics depends only on the electronic temperature of the normal metal \cite{Giazotto}. Therefore, the values of $V_{Tmeter}$ directly reflect $T_A$. 

The heater has been calibrated by acquiring the current-to-voltage characteristic of the tunnel junction at the different values of $T_{bath}$. In particular, the injected power is given by $P_{in}=0.5V_{heater}I_{heater}$, where the factor 0.5 stems from the fact that the heat is equally dissipated on the two sides of the junction. In this experimental configuration, the heat is generated at one side of the active region, while the thermometer is placed in the vicinity of the opposite end (see Figs. 1b and 3b). The electrons is $A$ thermalize through electron-electron interaction and electron-phonon scattering, while the heat losses through the thermometer tunnel junction is negligible compared to the other contributions (see Subsection \textit{Perfect Andreev mirrors}. As a consequence, the electronic temperature in the active region varies for distances of the order of the electron-phonon coherence length $l_{e-ph}=\sqrt{D_{Ave}\tau_{e-ph}}$ \cite{Giazotto}, where $D_{Ave}=(t_{Al}D_{Al}+t_{Cu}D_{Cu})/(t_{Al}+t_{Cu})\simeq5.6\times10^{-3}\text{m}^2/\text{s}$ is the diffusion constant of the active region and $\tau_{e-ph}=6\;\mu$s is the electron-phonon scattering time (see Section \textit{Bolometer} for details). Since the length of the active region is $L=1.5\;\mu$m$ \ll l_{e-ph}$, we can assume that the electronic temperature in the active region is homogeneous during our experiments.

\subsubsection{Perfect Andreev mirrors}
Figure \ref{fig_thermal1}-d shows $T_A$ as a function of the input power acquired at $T_{bath}=150$ mK in three different sets of measurements. The electronic temperature monotonically increases from 150 mK to $\sim270$ mK by rising $P_{in}$ to $\sim8$ fW.
Within the experimental parameters, the inequality $k_BT_A\ll \Delta_{0,S}$ is always satisfied. Therefore, the perfect mirror model (PMM) can describe the data: the heat exchange between $A$ and $S$s is fully suppressed ($P_{loss}=0$), i.e. \textit{A} is thermally isolated from $S$. Thus, the injected power $P_{in}$ relaxes only via electron-phonon interaction $P_{e-ph}$ and out-diffuses through the thermometer $IP$ $P_{AIP}$. The resulting quasi-equilibrium equation describing the PMM reads
\begin{equation}
P_{in}= P_{e-ph}+P_{AIP}\text{,}
\label{balance_equation}
\end{equation}
as schematically represented in Fig. \ref{fig_thermal1}-e.

  \begin{figure*}[htbp]
     \centering
     \includegraphics[width=1\textwidth]{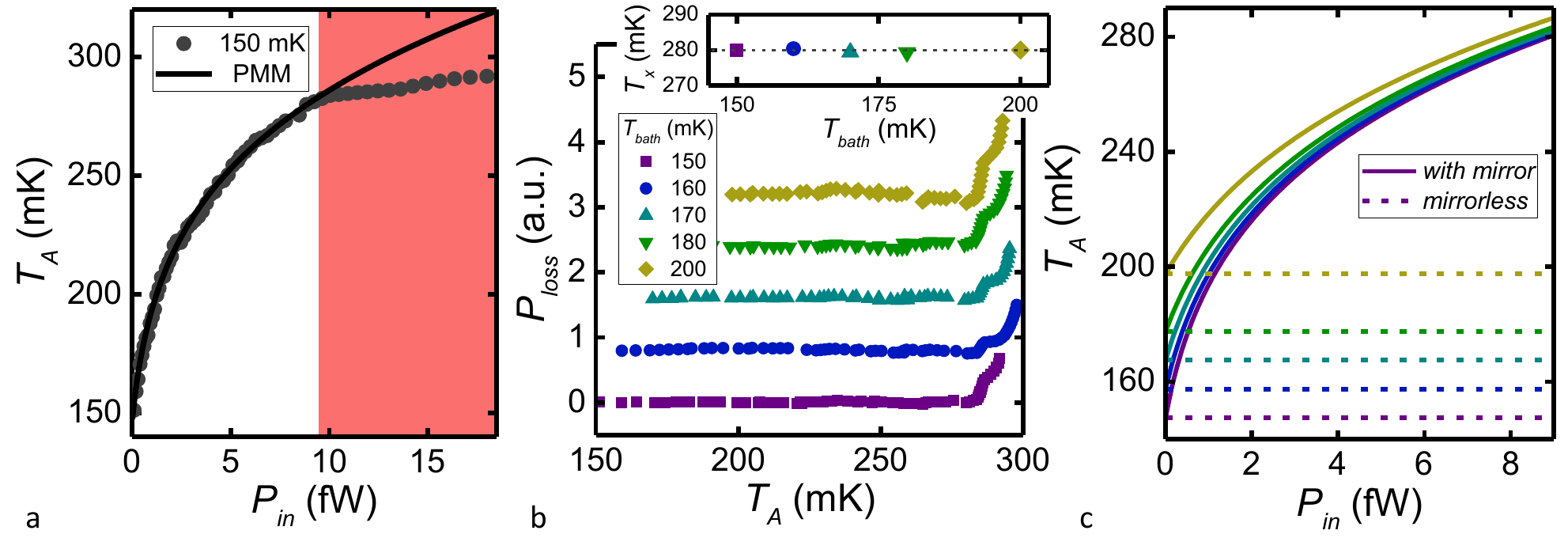}
     \caption{(a) Electronic temperature of the active region versus input power measured (grey points) and calculated with the PMM (black line) at $T_{bath}=150$ mK. The red shaded area highlights the values of $T_A$ where the PMM fails to describe the experimental data. (b) Power diffusion towards the superconducting leads $P_{loss}$ calculated as the difference between the experimental data and PMM theoretical curve at each $T_{A}$ for different values of $T_{bath}$. The curves have been vertically shifted for clarity. \textit{Inset}: Threshold temperature $T_{x}$ as a function of $T_{bath}$. $P_{loss}\neq 0$ for $T_A\geq 280$ mK independently from $T_{bath}$. (c) Calculated $T_A$ versus $P_{in}$ characteristics calculated in the presence (solid lines) and absence (dashed lines) of Andreev mirrors for different values of $T_{bath}$. The color code for $T_{bath}$ follows panel (c). The presence of Andreev mirrors is expected to strongly improve the power sensitivity of the nano-TES.}
     \label{fig_thermal2}
 \end{figure*}

Since $A$ is in the $\mathcal{N}$-state, the power exchanged via electron-phonon interaction can be written as \cite{Giazotto}:
\begin{equation}
P_{e-ph}= \Sigma_{A} \mathcal{V}_{A}\left(T_{A}^{5}-T_{bath}^{5}\right),
\label{e-phpower}
\end{equation}
where $\Sigma_{A}$ is the electron-phonon thermal relaxation constant and $\mathcal{V}_{A}$ is the volume of \textit{A}.
The power which flows through the thermometer JJ takes the form \cite{Giazotto}
\begin{align} 
P_{AIP}=\frac{1}{e^{2} R_{Tmeter}} \int_{-\infty}^{+\infty} \mathrm{d}E  E\;DOS_{P}(E,T_{bath})  \notag \\ 
\times \left[ f_{0}(E_A,T_{A})-f_{0}(E,T_{bath})\right],
\label{NISpower}
\end{align}
where $DOS_{P}(E,T_{bath})$ is the $DOS$ of the superconducting probe, $E_A=E-eV_{Tmeter}$ is the energy of the active region, and $f_{0}(E,T_{A,bath})=\left[1+\exp{\left(E/k_{B}T_{A,bath}\right)}\right]^{-1}$
is the Fermi-Dirac distribution of \textit{A} and \textit{P}, respectively.

By solving Eq. \eqref{balance_equation}, we fit the experimental electronic temperature of the $A$ as a function of $P_{in}$, as shown by the black line in Fig. \ref{fig_thermal1}-d. Since all the other device parameters are known ($\mathcal{V}_{A}= 38\times 10^{-22}$ m$^{3}$, $R_{Tmeter}=11.6$ k$\Omega$ and $T_{bath}=150$ mK), we extracted the value of the electron-phonon coupling constant of the Al/Cu bilayer $\Sigma_{A}\simeq 1.3\times 10^{9}$ W/m$^{3}$K$^{5}$. We notice that the PMM provides a remarkable fit of the experimental data thus describing correctly the system. The contribution of the electron-phonon relaxation is about one order of magnitude larger than the thermal losses through the thermometer tunnel junction. Therefore, the presence of the thermometer tunnel barrier as a negligible impact on the thermal experiment.
Furthermore, the resulting electron-phonon relaxation constant is in good agreement with the average of $\Sigma_{Cu}=2.0\times10^{9}$ W/m$^{3}$K$^{5}$ and $\Sigma_{Al}=0.2\times10^{9}$ W/m$^{3}$K$^{5}$ \cite{Giazotto}, weighted with the volumes of the copper and the aluminum layer forming the active region: $\Sigma_{A,theo}=(\Sigma_{Cu}\mathcal{V}_{Cu}+\Sigma_{Al}\mathcal{V}_{Al})/\mathcal{V}_{A}=1.38\times10^9$ W/m$^{3}$K$^{5}$, with $\mathcal{V}_{Al}\simeq1.58\times 10^{-21}$ m$^{-3}$ and $\mathcal{V}_{Cu}\simeq2.25\times 10^{-21}$ m$^{-3}$.

\subsubsection{Low-efficiency Andreev mirrors}

In order to test the AM efficiency and the limits of the PMM, we investigated the dependence of $T_A$ on larger values of $P_{in}$. 
The PMM fails for $P_{in}\ge9$ fW (the red shaded area in Fig. \ref{fig_thermal2}-a), where the power loss through the \textit{S}-electrodes is no longer negligible ($P_{loss}\neq 0$) and the resulting increase of $T_A$ is reduced.

The energy losses through the superconducting electrodes can be evaluated by calculating the difference between the measured $P_{in}$ necessary to produce a specific $T_A$ and its value estimated from the PMM. The dependence of $P_{loss}$ on $T_A$ is shown in Fig. \ref{fig_thermal2}-b for different values of $T_{bath}$. For all the curves, the energy loss through the Andreev mirrors is negligible until reaching a threshold temperature $T_{x}$. Notably, for all measurements we obtain $T_{x}\simeq 280$ mK (see the inset of Fig. \ref{fig_thermal2}-b), independently from the value of $T_{bath}$ and thus $T_S$. Furthermore, the energy filtering of the superconducting electrodes starts to fail for $T_{x}/T_{c,S}\sim 0.22$, that is in good agreement with the theoretical prediction of Andreev $\sim0.3T_c$ \cite{andreev}. Finally, the staircase behavior of $P_{loss}$ present at $T_A>280$ mK (see Fig. \ref{fig_thermal2}-b) could be due to superconducting proximity effect at the $A$/$S$ interface. However, a complete explanation of such behavior would require further measurements and analysis.

The $T_{A}$ versus $P_{in}$ characteristics change dramatically in a mirror-less device. Indeed, in a TES based on the same structure and dimensions but without AMs, the active region extends to the entire device, i.e. it is composed by a single superconductor. Fig. \ref{fig_thermal2}-c shows the difference between our nano-TES (solid lines) and a TES without Andreev mirrors (dashed lines) calculated for the same values of bath temperature of our experiments. As expected, at a given value of $P_{in}$ the temperature of the active region rises more in the presence of energy filtering than in a composite device, since the main channel for thermalization, the electron-phonon coupling, linearly depends on the volume (see Eq. \ref{e-phpower}). As a consequence, the presence of Andreev mirrors promises enhanced sensitivity of the nano-TES.

\section{Nano-TES Performance}
\label{Sec:Performance}

This section is devoted to the prediction of the performance of our device when operated as a radiation sensor. Our study will focus on both the bolometric operation, i.e. in continuous incident radiation, and the calorimetric operation, i.e. in single photon detection. Moreover, we propose a comparison between this device and an identical one without AM.

The typical read-out circuit for a TES is schematized in Fig. \ref{Fig:UnitCell}.
On the one hand, the decrease of the current $I_{TES}$ flowing through the inductance $L$ due to photon absorption can be measured by means of an inductively coupled superconducting quantum interference device (SQUID) amplifier.
On the other hand, the shunt resistor $R_{sh}$ implements the negative electro-thermal feedback mechanism (NETF), which guarantees constant voltage bias of the nano-TES and faster heat removal after radiation absorption \cite{IrwinBook}.
To this end, the shunt resistor needs to satisfy the relation $R_{sh} \ll R_A$ \cite{Irwin1995b}. In the following, we will use a value $R_{sh} = 10$ m$\mathrm{\Omega}$ typical for SQUID amplifier-based read-out.

\begin{table*}
\renewcommand\tabcolsep{11pt}
\begin{tabular}{c c c c c c c |c}
\noalign{\smallskip}\hline\noalign{\smallskip}

n-T & $\mathrm{T_c}$ & $\tau$ & $\tau_{eff}$ & $\mathrm{NEP_{TFN}}$ & $\mathrm{NEP_{tot}}$ & $\delta\nu$ &  $\nu/\delta\nu$ \\ 

& (mK) & ($\mu s$) & ($\mu s$) & (W/$\mathrm{\sqrt{Hz}}$)& (W/$\mathrm{\sqrt{Hz}}$) & (GHz) &   \\  

&&&&&&& 100 GHz \quad 300 GHz  \quad 1 THz\\

\noalign{\smallskip}\hline\noalign{\smallskip}
1 & 128 & 6 & 0.01 & 5.2 x $10^{-20}$ & 5.2 x $10^{-20}$     & 100 &  1 \qquad\qquad 3 \qquad\qquad 10 \\

1*&& $6$ & $0.01$ & 1.1 x $10^{-16}$& 4.7 x $10^{-16}$& 2 x $10^5$&  \quad 4 x $10^{-4}$ \quad 1 x $10^{-3}$ \quad 4 x $10^{-3}$\\

2 & 139 & 5 & 0.2 & 6.7 x $10^{-20}$ & 6.7 x $10^{-20}$ & 540 &  0.18 \qquad\quad 0.55 \quad\qquad 1.8 \\

2*&&$5$ & $0.2$& 1.5 x $10^{-16}$& 8.3 x $10^{-15}$& 1 x $10^6$&  \quad 8 x $10^{-5}$ \quad  2 x $10^{-4}$ \quad  8 x $10^{-4}$\\

\noalign{\smallskip}\hline

\end{tabular}
\caption{\textbf{Principal figures of merit.} The time constant $\tau$, the pulse recovery time $\tau_{eff}$, the Noise Equivalent Power due to the thermal fluctuation noise ${NEP_{TFN}}$ and the total noise ${NEP_{tot}}$, the Frequency Resolution $\delta \nu$,  and the Resolving Power $\nu/\delta\nu$ (at $100-300-1000$ GHz) are reported for two fabricated nano-TESs (n-T1 and n-T2), which can operate around own critical temperature ${T_c}$. The n-T1* and n-T2* values are referred to the superconducting elements without the Andreev mirrors.}
\label{tab:FigureOfMerit}
\end{table*}

\subsection{Bolometer}
Starting from the structure and the measured parameters, we evaluate the performance of our nano-TESs in terms of response time $\tau$ and $NEP$. The parameters that we will deduced in this section are reported in Tab. \ref{tab:FigureOfMerit}.

The thermal response time $\tau$ defines the dissipation rate of the overheating arising from radiation absorption in $A$. The value of $\tau$ is related to the quasi-particle thermalization with the phonons residing at $T_{bath}$. Namely, it depends on the electron heat capacity $C_{e,A}$ and the thermal conductance $G_{th,A}$ of $A$ through \cite{Irwin1995a}
\begin{equation}
    \tau=\frac{C_{e,A}}{G_{th,A}}\text{.}
    \label{time}
\end{equation}

The $A$ electron heat capacitance reads
\begin{equation}
    C_{e,A}=\Upsilon_{A}\mathcal{V}_{A} T_{c,A}\text{,}
    \label{thermalC}
\end{equation}
with $\Upsilon_{A}$ the Sommerfeld coefficient of $A$. Since $A$ is formed by an Al/Cu bilayer, we have to substitute  $\Upsilon_{A}\mathcal{V}_{A}=\Upsilon_{Cu}\mathcal{V}_{Cu}+\Upsilon_{Al}\mathcal{V}_{Al}$ (with $\Upsilon_{Cu}=70.5$ JK$^{-2}$m$^{-3}$, $\Upsilon_{Cu}=91$ JK$^{-2}$m$^{-3}$) in Eq. \ref{thermalC} .

The $A$ total thermal conductance $G_{th,A}$ is the derivative of the heat losses of $A$ with respect to its electronic temperature \cite{Giazotto,Irwin1995a}. Considering the nano-TES operation at $T_{c,A}\ll T_{c,S}$ we can consider $P_{loss}=0$, thus electron-phonon relaxation is the only relevant thermal exchange mechanism. Therefore, the total thermal conductance reads
\begin{equation}
    G_{th,A}=\frac{dP_{e-ph}}{dT_{A}}=5 \Sigma_{A} \mathcal{V}_{A} T^{4}_{A}\text{.}
    \label{thermalG}
\end{equation}

In our nano-TESs, the electron heat capacitance is ${C_{e,A}}_1 =4 \times 10^{-20}$ J/K for n-T1 (${C_{e,A}}_2 = 4.2 \times 10^{-20}$ J/K for n-T2), while the thermal conductance is as low as ${G_{th,A}}_1= 6.7\times10^{-15}$ W/K (${G_{th,A}}_2=9.3\times 10^{-15}$ W/K). As a consequence, the active region relaxation time is limited by $G_{th,A}$ to a few microseconds for both devices ($\tau_1\simeq6\;\mu$s and $\tau_2\simeq5\;\mu$s).

\begin{figure}[t!]
    \includegraphics[width=0.5\textwidth]{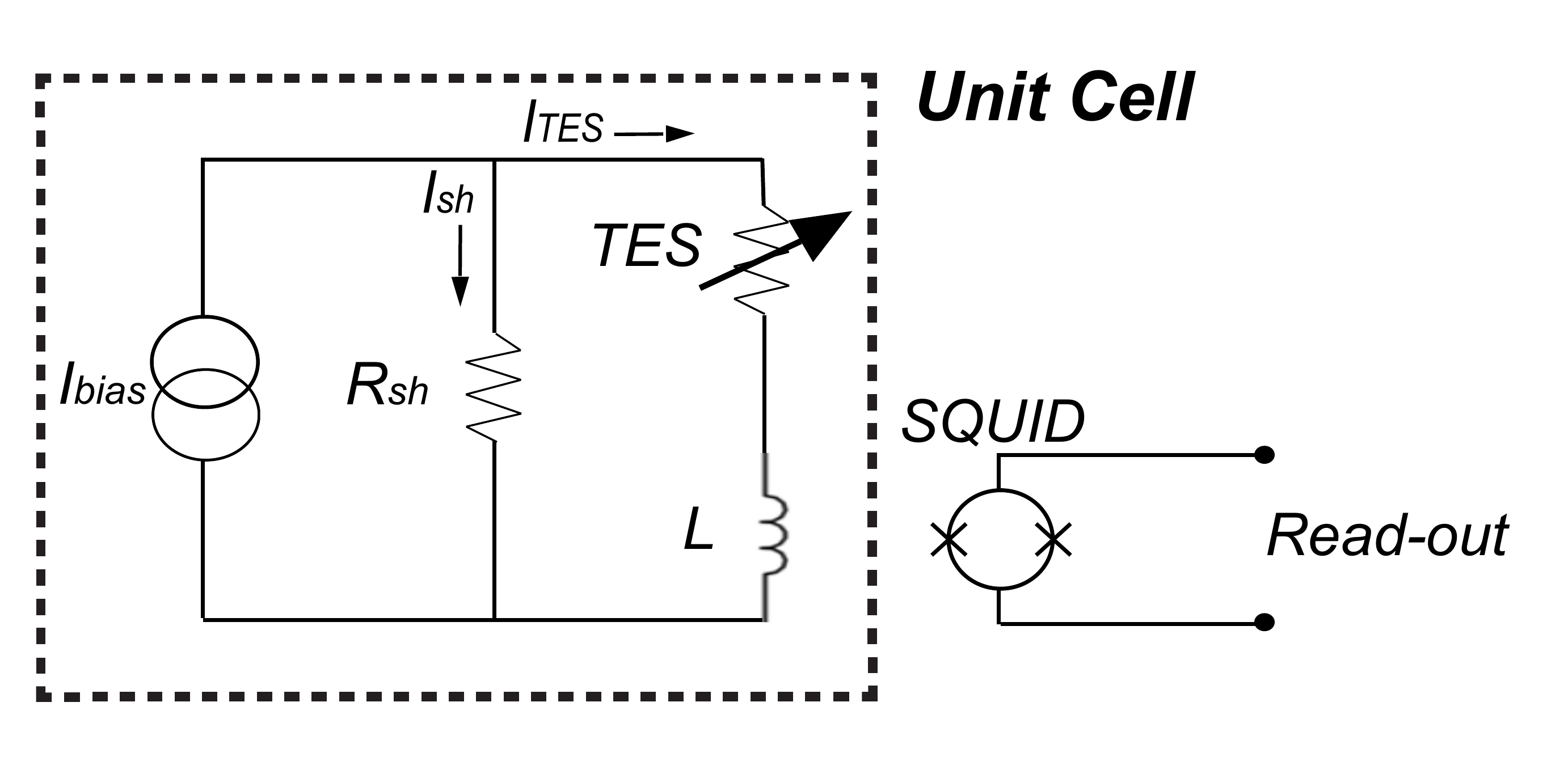}
    \caption{Schematic of the read-out circuit of a single nano-TES. The dashed line highlight the circuit providing the negative electro-thermal feedback NETF, where the nano-TES is biased by the current $I_{TES}$ and operates in parallel to the shunt resistor $R_{sh}$. The $I_{TES}$ variations are measured with a SQUID amplifier coupled to the circuit by an inductance $L$.}
    \label{Fig:UnitCell}
\end{figure}

The negative electro-thermal feedback (see Fig. \ref{Fig:UnitCell}) affects the thermal response of the nano-TES. In particular, the sharpness of the superconducting to normal-state phase transition defines the effective recovery time through $\tau_{eff} = \frac{\tau}{1 + \alpha/n}$, where typically $n=5$ for a clean metal and $\alpha = \frac{d\log(R)}{d\log(T)}$ is the electro-thermal parameter that takes into account the sharpness of the transition from the superconducting to the normal-state.  As reported in Tab. \ref{tab:FigureOfMerit}, $\tau_{eff}$ is one or two orders of magnitude smaller than the thermal response time ($\tau_{eff} << \tau$). Namely, the effective response time is $\tau_{eff,1} = 0.01\;\mu$s and $\tau_{eff,2} = 0.2\;\mu$s for n-T1 and n-T2, respectively.

The $NEP$ is the most important figure of merit for a bolometer, since it determines the minimum power that can be detected above the noise level. Taking into account the equivalent circuit, highlighted by the dashed line in Fig. \ref{Fig:UnitCell}, the total $NEP$ of the nano-TES is given by three uncorrelated sources \cite{Mather,Lee} 
\begin{equation}
    NEP_{tot} \simeq \sqrt{NEP^2_{TFN} + NEP^2_{Jo} + NEP^2_{sh} }\text{,}
    \label{Eq:totalNEP}
\end{equation}
where ${NEP_{TFN}}$ is associated to the thermal fluctuations, ${NEP_{Jo}}$ is due to the Johnson noise in the nano-TES and ${NEP_{sh}}$ is related to the shunt resistor. Other external noise contributions, such as the noise of the read-out electronics and the photon background noise, are not taken into account, because they can not be directly attributed to the device.

The thermal fluctuation noise given by \cite{Bergmann}
 \begin{equation}
      NEP_{TFN} = \sqrt{4 k_B \Lambda G_{th,A} {T_{c,A}}^{2}}\text{,} \\
      \label{Eq:NEP_TFN}
 \end{equation}
where $\Lambda = n /(2n+1)$ describes the effect of the temperature gradient across the thermal link. Our nano-TESs show an extremely low  thermal fluctuation noise, $NEP_{TFN,1}=5.2\times10^{-20}$ W/$\mathrm{\sqrt{Hz}}$ and $NEP_{TFN,2}=6.7\times10^{-20}$ W/$\mathrm{\sqrt{Hz}}$, since $G_{th}$ is limited by the small volume of the active region.

The Johnson noise is originated by the charge transport, when the nano-TES is in the normal-state. The related \textit{NEP} is written \cite{Bergmann}
\begin{equation}
    NEP_{Jo}  = \sqrt{4 k_B R_A(T_c) T_{c,A}} \frac{G_{th,A} T_{c,A}}{V \alpha} \sqrt{1+ 4\pi^2 f^2 \tau_{eff}^2} \text{,}
    \label{Eq:NEP_Jo}
\end{equation}
where $R_A(T_c) \simeq$ 40 $\Omega$  is the value of $R_A$ at $T_c$, $V$ is the voltage drop and $f$ is the signal bandwidth. In order to detect the temperature variations in $A$, we chose a signal bandwidth $f_1=100\;\text{MHz}\geq 1/\tau_{eff,1}$ and $f_2=5\;\text{MHz}\geq1/\tau_{eff,2}$ for n-T1 and n-T2, respectively. For $I_{TES}=15$ nA, our devices show similar normal state resistances ($R_{A,1}\simeq 70\;\Omega$ and $R_{A,2}\simeq 72\;\Omega$), but different values of the electro-thermal parameter ($\alpha_1=2742 $ ~and $\alpha_2=122$). Therefore, the Johnson contributions to the noise equivalent power are $NEP_{Jo,1}=6\times10^{-23}$ W/$\mathrm{\sqrt{Hz}}$ and $NEP_{Jo,2}=2\times10^{-21}$ W/$\mathrm{\sqrt{Hz}}$ for n-T1 and n-T2, respectively.

Finally, the shunt noise is related to charge fluctuations through $R_{sh}$. Its contribution to the $NEP$ reads \cite{Bergmann}
     \begin{align}
            NEP_{sh}  = \sqrt{4 k_B R_{sh}  T_{bath}} &\frac{G_{th,A} T_{c,A}}{V \alpha} \notag \\      \times  &\sqrt{(1-L_0)^2 + 4\pi^2 f^2 \tau_{eff}^2}, 
       \label{Eq:NEP_Sh}
       \end{align}
where $L_0 = \alpha/n$ is the loop gain. Since the shunting resistor needs to satisfy $R_{sh}\ll R_A$, the contribution of $NEP_{sh}$ is usually negligible compared to Johnson noise. Indeed, in our nano-TESs we have $NEP_{sh,1}=3\times10^{-24}$ W/$\mathrm{\sqrt{Hz}}$ and $NEP_{sh,2}=1\times10^{-22}$ W/$\mathrm{\sqrt{Hz}}$ for n-T1 and n-T2, respectively.

The total noise equivalent power of our nano-TESs is dominated by the thermal fluctuation contribution, that is Johnson and shunt resistor noise are negligible, and it shows state-of-the-art values $NEP_{tot,1}=5.2\times10^{-20}$ W/$\mathrm{\sqrt{Hz}}$ ~and $NEP_{tot,2}=6.7\times10^{-20}$ W/$\mathrm{\sqrt{Hz}}$ for TES technology \cite{Morgan}.

A full analysis of nano-TES performance as bolometer requires an evaluation of the dynamic range in response to large signals. The saturation power in the limit of voltage bias and narrow superconducting-to-normal-state transition can be written as \cite{IrwinBook}
\begin{equation}
P_{sat} = \left(1- \frac{R_A(T_c)}{R_A}\right)P_{e-ph},
\end{equation}
where $R_A(T_c)$ $\simeq40$ $\Omega$ is the resistance of the active region at $T_c$, $R_A\simeq70$ $\Omega$ is the normal-state resistance. The saturation power ranges approximately from 50 to 100 aW for both devices, thus confirming the predicted high extremely sensitivity of our structures when operated as nano-TESs. These values are promising and well suited in medical, industrial and astronomical applications. It is anyways possible to increase the saturation power by increasing the heat losses through the phonons, that is by increasing the active region volume, with the simultaneous increase of the $NEP$.

We now consider devices characterized by the same structure of our nano-TESs, but fabricated without the lateral aluminum banks, namely they are completely made of the Al/Cu bilayer. The result of this structure is the absence of heat confinement in the small nanowire and the increase of the net device volume to about $1.8\times10^{-14}$ m$^3$. On the one hand, this change of structure does not affect the thermal response time, since both the heat capacity and the thermal conductance depend linearly on the volume (see Eqs. \ref{thermalC} and \ref{thermalG}, respectively). On the other hand, the total noise equivalent power is strongly influenced by the volume increase ($NEP_{tot,1^*}=5\times10^{-16}$ W/$\mathrm{\sqrt{Hz}}$ and $NEP_{tot,2^*}=8\times10^{-15}$ W/$\mathrm{\sqrt{Hz}}$). In particular, $NEP_{Jo}$ and $NEP_{sh}$ depend linearly on $G_{th}$ thus showing the larger worsening ($NEP_{Jo,1^*}=4\times10^{-16}$ W/$\mathrm{\sqrt{Hz}}$,~ $NEP_{Jo,2^*}=5\times10^{-16}$ W/$\mathrm{\sqrt{Hz}}$ and $NEP_{sh,1^*}=4\times10^{-16}$ W/$\mathrm{\sqrt{Hz}}$, ~ $NEP_{sh,2^*}=6\times10^{-16}$ W/$\mathrm{\sqrt{Hz}}$) and becoming sizeable with respect to thermal fluctuations ($NEP_{TFN,1^*}=1\times10^{-16}$ W/$\mathrm{\sqrt{Hz}}$ ~and $NEP_{TFN,2^*}=1.4\times10^{-16}$ W/$\mathrm{\sqrt{Hz}}$). Therefore, the removal of AM has a heavy negative impact on the detection performance of the TES bolometer.

\subsection{Calorimeter}
In single-photon detection, the value of $\tau_{eff}$ determines the minimum speed of the read-out electronics necessary to detect a single photon. Moreover, it defines the dead time, that is the minimum time interval between two incoming photons in order to be recorded as two different events. The NETF ensures that the energy injected into the sensor by the single photon absorption is efficiently removed by decreasing its Joule overheating instead of being dissipated through the substrate thus compensating for the initial temperature increase.

The fundamental figure of merit for a single-photon detector is the frequency resolution $\delta \nu$, that is the minimum photon frequency detected by the sensor. For the nano-TES, it is defined \cite{Irwin1995a}
\begin{equation}
\delta\nu = \dfrac{2.36}{h} \sqrt{4\sqrt{\dfrac{n}{2}}k_B T_c^2 \dfrac{C_{e,A}}{\alpha}}\text{.}
\label{Eq:FreqResol} 
\end{equation}

\begin{figure}
    \includegraphics[width=0.49\textwidth]{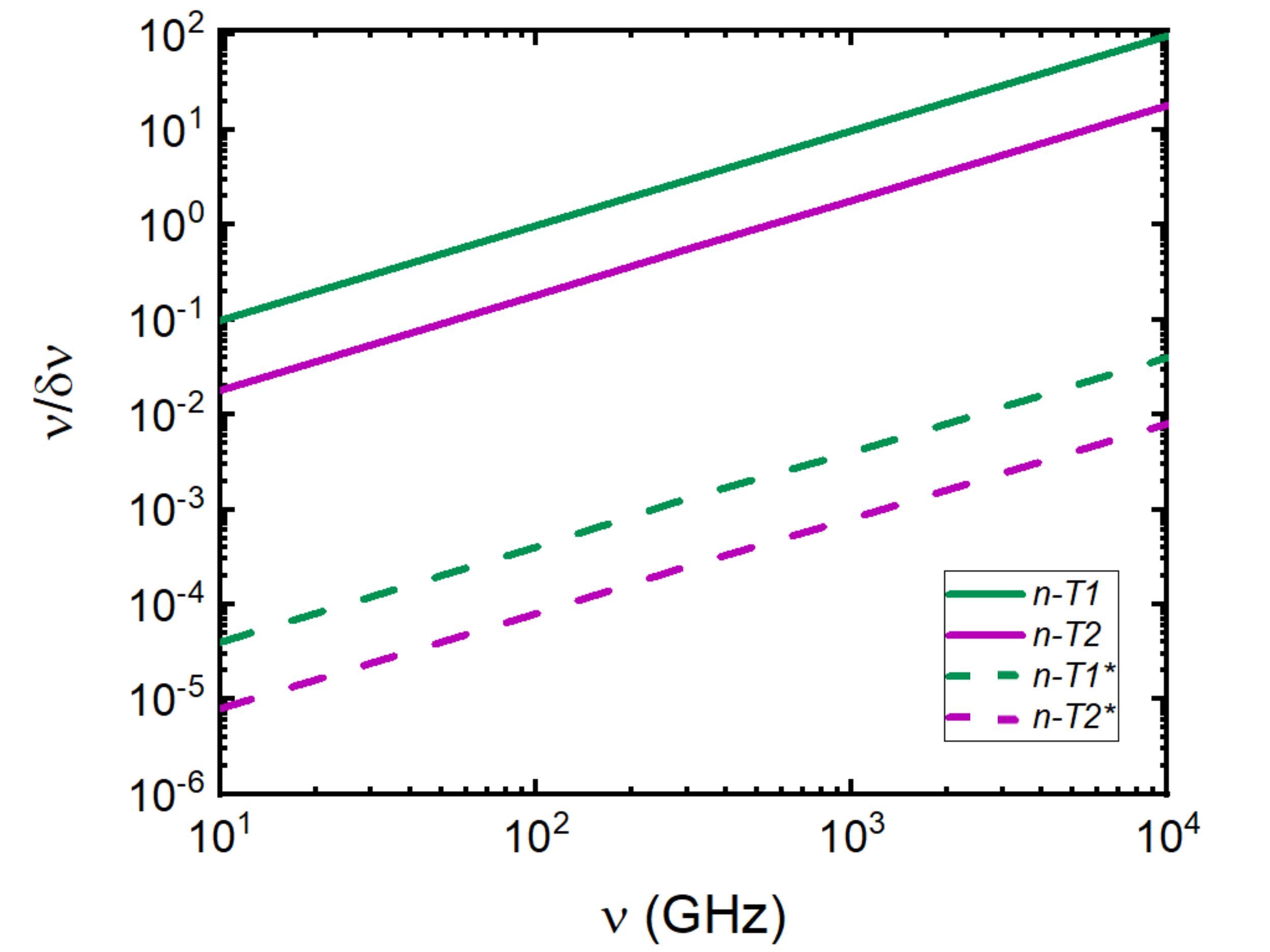}
    \caption{Resolving power as a function of single-photon frequency for n-T1 (green) and n-T2 (purple). The dashed lines represent similar devices not equipped with Andreev mirrors. The presence of Andreev mirrors improved the resolving power of more than 3 orders of magnitude.}
    \label{Fig:ResolvingPower}
\end{figure}

Since the minimum detectable single-photon energy depends on $\alpha$, our nano-TESs show different values of $\delta\nu$. In particular, we have $\delta\nu_1\simeq$ 100 GHz ($\delta E_1\simeq$ 0.4 meV) and $\delta\nu_2\simeq$ 540 GHz ($\delta E_2 \simeq$ 2 meV) for n-T1 and n-T2, respectively. Accordingly, the resolving power ($\nu/\delta \nu$), which indicates the sensitivity in detecting radiation of a specific energy, achieves values larger than $1$ for $\nu \geq100$ GHz for n-T1, as shown in Fig. \ref{Fig:ResolvingPower}. We note that the sensitivity of the devices could be further improved by increasing the sharpness of the superconducting to normal-state transition of the active region, i.e. rising the value of $\alpha$.

In the absence of AM, the electron heat capacitance increases of about 7 orders of magnitude due the volume increase. Therefore, the frequency resolution downgrades of more than 3 orders of magnitude (see Tab. \ref{tab:FigureOfMerit}), and the devices could operate only above 300 THz.

\section{Multiplexing circuits for detector arrays}
\label{Sec:Readouts}
Astronomy and astrophysical experiments require telescopes equipped with arrays of hundreds or thousands detectors. Therefore, efficient multiplexing schemes are fundamental to decrease the wiring, lower the related noise, and reduce the mechanical and thermal loads. Several multiplexing architectures differing for the output signals are used: time division multiplexing TDM, code division multiplexing CDM, frequency division multiplexing FDM and microwave resonator MR based read-out \cite{Ullom}.

For the frequency operation range and the target applications of our nano-TESs, FDM and MR represent the optimal strategies to create multipixel detectors. Thus, we will estimate the circuit parameters to build arrays of our nano-TESs.

\begin{figure}
    \includegraphics[width=\columnwidth]{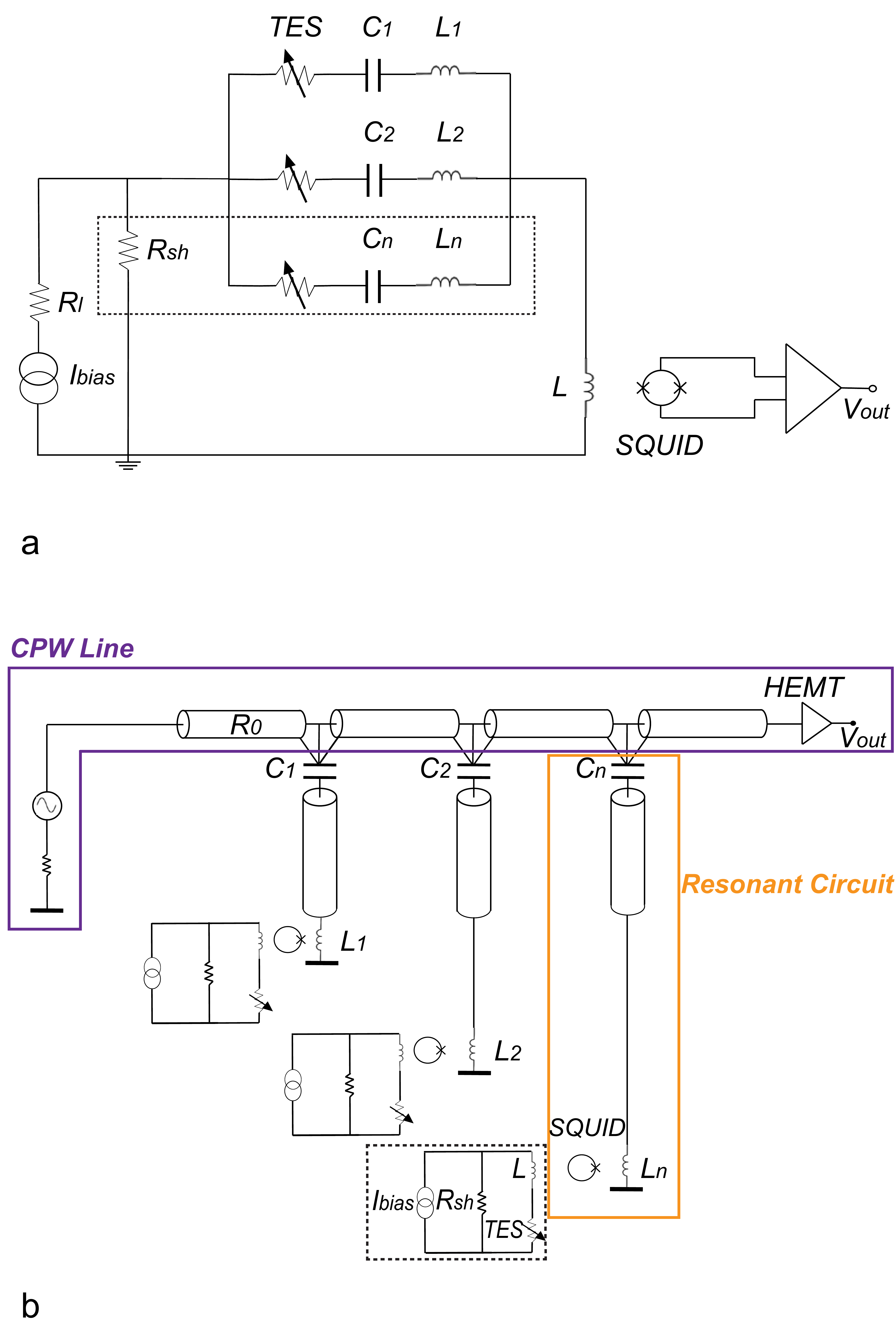}
    \caption{\textbf{Multiplexing read-out circuits.} a) Scheme of frequency division multiplexing (FDM) read-out. FDM is composed by N independent unit cells, which resonate at different frequencies set by the RLC circuit (${L_n}= 10 ~\mu$H, a capacitance range $C$ between $10$ pF and $25$ pF, ${R_A}(T_c)$ $= 40~{\Omega}$). Their frequencies are summed and coupled with a SQUID linked to an amplifier. b) Scheme of microwave resonator (MR) based multiplexing read-out. The schematic circuit is formed by N independent unit cells coupled with an RC resonant circuit, composed by an inductance $L$, a coaxial cable and a capacitance $C$, through a SQUID determining a unique frequency for each resonator. All resonators are read out by a frequency comb applied by an AC generator through a Coplanar Waveguide (CPW) Line. The signal can be amplified by an High Electron Mobility Transistor (HEMT) (${R_0} = 50~ {\Omega}$, ${C_n}$ between $60$ pF and $2$ nF.}
    \label{fig:FDM_MWRes}
\end{figure}

\subsection{Frequency Division Multiplexing}
\label{Sub:FDM}
The FDM circuit is schematically shown in Fig \ref{fig:FDM_MWRes}-a \cite{Lanting,Vaccaro}. Each unit cell operates at its own frequency [$f_n=1/(2\pi\sqrt{L_n C_n})$] defined by the RCL circuit and adequately separated from the others to avoid cross talk.
The signal bandwidth of each pixel is larger than the relaxation effective time-scale of the nano-TES after the photon absorption ($BW  > 1/ \tau_{eff}$) suppressing all noise signals outside the band.
In order to have a signal bandwidth [$BW = {R_A}(T_c) /(2\pi L_n)$] constant for each pixel of the array, the same inductance is usually set for every channel. For example, considering the values of n-T2 we propose a bandwidth $BW$ of $60$ kHz, an inductance of $10$ $\mu$H and a capacitance range between $10$ pF and $25$ pF to have $\sim 31$ pixels for an array. Therefore, each pixel has a resonance frequency between $10$ MHz and $16$ MHz, and it is spaced of $200$ kHz to suppress cross-talk.
The total signal is measured with a single SQUID amplifier of time constant shorter than the effective pulse recovery time to follow the current variation in the nano-TES ($\mathrm{\tau_s \ll \tau_{eff}}$). Its bandwidth BW of 20 MHz is consistent with state-of-the-art SQUID amplifiers \cite{Muck,Huber}. Custom LC lithographed boards with multiplexing factors $>50$ can be fabricated to this end. The generation of tones and subsequent demodulation can be handled by electronic boards equipped with FPGAs suitably designed \cite{Dobbs}.

\subsection{Microwave resonator multiplexing}
\label{Sec:MWResonator}
The MR multiplexing exploits a SQUID amplifier connecting the sensing elements of each pixel to a different RLC resonant circuit (see Fig. \ref{fig:FDM_MWRes}-b). This configuration maximizes the dynamic range per pixel and removes the limit on the pixel number, but the system is more bulky.\\
Photon absorption shifts the resonance of the related circuit which is connected in parallel and excited simultaneously to all the others. Then, the transmitted signals are summed into a low noise amplifier, such as the high electron mobility transistor (HEMT) placed at a higher temperature.
The bandwidth of n-T2 is $BW = 5$ MHz. Considering a SQUID amplifier with bandwidth $BW_s= 10$ MHz, we can choose resonant circuits with bandwidth $BW_r= 50$ MHz  separated by $50$ MHz in frequency range going from $300$ MHz to $2$ GHz.
To this end, each line could implement systems with resistance of $\mathrm{50 ~\Omega}$, fixed inductance of 0.2 $\mathrm{\mu}$F and capacitance values between $0.04$ pF and $1.4$ pF. Therefore, the MR would allow to build arrays of nano-TESs with high dynamic range, low power dissipation and giving the possibility to select a larger bandwidth.

\section{Summary and conclusions}
We have presented an extra-sensitive and optimized miniaturized structure, which can be used as a nanoscale transition edge sensor (nano-TES). The ultra-low volume of the active region and the exploitation of heat barriers, the so-called Andreev mirrors, ensure the optimal thermal efficiency of the devices. In addition, the engineering of the working temperature thanks to the superconducting inverse proximity effect allows full control of the nano-TES performance. To extract all the device parameters and determine the performance both in the bolometer and calorimeter operation, we performed a complete series of experiments. On the one hand, by characterizing electrically the nano-TES we measured the critical current and the critical temperature of the active region. On the other hand, we fabricated and characterized a secondary device equipped with superconducting tunnel probes extracting the spectral and thermal properties of $A$. 

Starting from the experimental data, we calculated the performance of our device when operated as a nano-TES by employing widespread and well known equations. The nano-TES is predicted to reach a total noise equivalent power of $\sim 5 \times 10^{-20}$ W/$\sqrt{\text{Hz}}$, limited exclusively by the thermal fluctuations, when operated as bolometer. In single-photon detection, our device shows a best frequency resolution of $\sim100$ GHz, thus having the potential to operate in THz and sub-THz regime with a relaxation time of $\sim$ 10 ns.

With its simple design, the nano-TES could be implemented in widespread multiplexing circuits (FDM and MW Resonators) for detector arrays in multipixel gigahertz cameras \cite{Ullom}. As a consequence, the nano-TES could become an asset as bolometer and calorimeter for astronomy and astrophysics research, detecting cosmic microwave background \cite{Sironi} and atomic vibration in galaxy cluster \cite{Villaescusa-Navarro}, and be the key for searching axions \citep{Redondo, Ringwald, Spagnolo}, one of the principal candidates of dark matter. Moreover, it could find application for medical imaging \cite{Sun}, industrial quality controls \cite{Ellrich} and security \cite{Rogalski} in THz band.

\section*{acknowledgements}
We acknowledge A. Tartari, and G. Lamanna for fruitful discussions.
The authors acknowledge  the European Union's Horizon 2020 research and innovation programme under the grant No. 777222 ATTRACT (Project T-CONVERSE) and under grant agreement No. 800923-SUPERTED. The authors acknowledge CSN V of INFN under the technology innovation grant SIMP. The work of F.P. was partially supported by the Tuscany Government (Grant No. POR FSE 2014-2020) through the INFN-RT2 172800 project. The work of V.B. is partially funded by the European Union (Grant No. 777222 ATTRACT) through the T-CONVERSE project. G.S. acknowledges the ASI grant 2016-24-H.0.

%
\section*{Conflict of interest}
The authors declare that they have no conflict of interest.

\section*{Data availability}
The data that support the findings of this study are available from the corresponding author upon reasonable request.


\end{document}